\documentclass[12pt,preprint]{aastex}

\usepackage{psfig}

\shorttitle{Variability of the Radio Emission from PSR J0529$-$6652}
\shortauthors{Crawford et al.} 
 
\begin{document}

\title{Variability of the Pulsed Radio Emission from the Large
Magellanic Cloud Pulsar PSR J0529$-$6652}

\author{F. Crawford\altaffilmark{1}, D. Altemose\altaffilmark{1}, 
H. Li\altaffilmark{1}, \& D. R. Lorimer\altaffilmark{2}}

\altaffiltext{1}{Department of Physics and Astronomy, Franklin \& 
Marshall College, P.O. Box 3003, Lancaster, PA 17604, USA}

\altaffiltext{2}{Department of Physics, West Virginia University,
Morgantown, WV 26506, USA}
 
\begin{abstract}
We have studied the variability of PSR J0529$-$6652, a radio pulsar in
the LMC, using observations conducted at 1390 MHz with the Parkes 64-m
telescope. PSR J0529$-$6652 is detectable as a single pulse emitter,
with amplitudes that classify the pulses as giant pulses. This makes
PSR J0529$-$6652 the second known giant pulse emitter in the LMC,
after PSR B0540$-$69.  The fraction of the emitted pulses detectable
from PSR J0529$-$6652 at this frequency is roughly two orders of
magnitude greater than it is for either PSR B0540$-$69 or the Crab
pulsar (if the latter were located in the LMC). We have measured a
pulse nulling fraction of $83.3 \pm 1.5$\% and an intrinsic modulation
index of $4.07 \pm 0.29$ for PSR J0529$-$6652.  The modulation index
is significantly larger than values previously measured for typical
radio pulsars but is comparable to values reported for members of
several other neutron star classes.  The large modulation index, giant
pulses, and large nulling fraction suggest that this pulsar is
phenomenologically more similar to these other, more variable sources,
despite having spin and physical characteristics that are typical of
the unrecycled radio pulsar population. The large modulation index
also does not appear to be consistent with the small value predicted
for this pulsar by a model of polar cap emission outlined by
\citet{gs00}.  This conclusion depends to some extent on the
assumption that PSR J0529$-$6652 is exhibiting core emission, as
suggested by its simple profile morphology, narrow profile width, and
previously measured profile polarization characteristics.
\end{abstract}


\section{Introduction}

From the time of their discovery more than 40 years ago, the detailed
mechanism by which pulsars generate radio emission has remained
elusive. Models have been proposed over that time to explain pulsar
emission, and in principle these models can be tested with pulsar
observations. One relatively recent emission model, proposed by
\citet{gs00}, is an extension of the polar cap spark model of
\citet{rs75} in which bursts of plasma (``sparks'') are produced at
the polar cap. In this model, instabilities in the plasma traveling
along the magnetic field lines induce radio emission. The model
postulates a densely packed region of these sparks, with a
characteristic size and separation $h$. One can determine a
characteristic number of sparks $a$ (in one dimension) across the
polar cap according to

\begin{equation}
a = r_{p} / h,
\end{equation}

where $r_{p}$ is the polar cap radius. In the \citet{gs00} model, $a$
is a ``complexity parameter'' that can also be determined
observationally by

\begin{equation}
a = 5 (\dot{P} / 10^{-15})^{2/7} (P/1~{\rm s})^{-9/14},
\end{equation}

where $P$ and $\dot{P}$ are the pulsar's spin period and period
derivative, respectively.  There are several other similar models
having different scalings to determine the complexity parameter (see,
e.g., \citet{bjb+12} for a brief summary and references to these
models).  \citet{jg03} provide details on how such models might be
tested by comparing a pulsar's observed intensity modulation with the
variability predicted by the complexity parameter in the model.
Although the \citet{gs00} model does not make a distinction between
core and conal emission from the pulsar \citep{r83}, a model test is
most useful using pulsars with core emission, since observationally it
is the most straightforward probe of the polar cap region, it is not
affected by viewing geometry effects from an off-center cut through
the emission region, and it is less likely to exhibit cone-based pulse
drifting effects which would increase the observed modulation
\citep{wes06}.

One measure of a pulsar's intensity variability that can be employed
in such tests is the phase-resolved modulation index, $m(\phi)$. Here
$\phi$ is the longitude of the pulse phase, and $m(\phi)$ is the ratio
of the standard deviation $\sigma(\phi)$ of the observed pulse
intensities to the mean pulse intensity averaged over many adjacent
pulses, $\langle I(\phi) \rangle$ \citep{jg03, wes06}:

\begin{equation}
m(\phi) = \frac{\sigma(\phi)}{\langle I(\phi) \rangle} =
\frac{\sqrt{\langle I(\phi)^{2} \rangle - \langle I(\phi)
\rangle^{2}}}{\langle I(\phi) \rangle}.
\label{eqn-2}
\end{equation}

As outlined in \citet{jg03}, one would expect an anti-correlation
between $a$ and $m$ owing to the likelihood that overlapping sparks in
a complex emission region (large $a$) would wash out and tend to
reduce the degree of modulation, creating more steady emission
(smaller $m$).  The phase information is important to retain since the
measured modulation index near the edges of a profile peak usually
increases relative to the center, in some cases because conal emission
may be present.  The center of the peak of the pulse profile therefore
offers the best measurement of the properties of the core
emission. The largest signal is also present at the profile peak, so
the most precise constraint on $m(\phi)$ can generally be made at that
pulse phase.

The modulation index has previously been measured for a number of
radio pulsars. \citet{wes06} observed 187 pulsars at 1400 MHz to
investigate the modulation and sub-pulse drifting properties of the
sample, and they produced modulation index measurements for 175 of
them.  More recently, \citet{bjb+12} reported measurements for a
sample of 103 pulsars from the HTRU 1400-MHz survey \citep{kjv+10}.
\citet{jg03} analyzed 2 pulsars observed near 1400 MHz and re-analyzed
another 10 that were previously observed at 430 MHz \citep{jap01,
wab+86}. The \citet{jg03} sample includes a subset of 28 pulsars
analyzed by \citet{wab+86} at 430 MHz. \citet{jg04} also measured $m$
for PSR B1937+21 at 1400 MHz. Other measurements of $m$ (but not phase
resolved) were reported in Table 4 of \citet{wje11} for members of
several neutron star classes having different types of emission. These
objects were the Crab pulsar -- a classical giant pulse emitter
\citep{sr68, ag72}, the radio-emitting magnetar XTE J1810$-$197
\citep{crh+06}, the rotating radio transient (RRAT) J1819$-$1458
\citep{mll+06}, and the possible RRAT-link pulsar PSR B0656+14
\citep{wsr+06}. The modulation indices reported for three of these
four sources were for frequencies near 1400 MHz; PSR B0656+14 was
measured at 327 MHz.

Generally, $m$ is observed to be small, with typical values between
about 0.5 and 1 \citep{wes06, wje11, bjb+12}.  The histogram of
measured modulation indices shown in Fig. 8 of \citet{wes06} shows no
measured $m$ with a value above 2, and only two of the 103
measurements of \citet{bjb+12} have $m > 2$.  As seen in Fig. 1 of
\citet{jg03} and Fig. 10 of \citet{wes06}, the anti-correlation
between $m$ and $a$ in the observed sample of pulsars is weak.  This
is also seen in the \citet{bjb+12} sample. One reason may be the
inclusion of pulsars with conal emission in these larger samples; the
modulation index and complexity parameter are not predicted to be as
strongly correlated for conal emitters as they are for pulsars with
core emission \citep{jg03, bjb+12}. Another difficulty in establishing
correlations is that the range of $m$ measured in the observed data
set is small.  One well-known exception to this is the Crab
pulsar. The Crab's modulation index is due to its well-studied giant
pulses \citep{sr68, ag72, lcu+95, wes06, ksv10}.  However, this pulsar
is somewhat atypical since it is young and very energetic and is also
a well-known X-ray emitter. In addition, the Crab pulsar is difficult
to classify as a core or conal emitter \citep{r90}. The Crab may have
conal emission and other complicating geometric profile features that
could produce a larger measured modulation index, making its use for a
simple test of the \citet{gs00} model difficult.

However, PSR J0529$-$6652 is one pulsar that could be used for a
consistency check of the \citet{gs00} model. This is a 976-ms pulsar
that was first discovered in a 600 MHz pulsar survey of the Large
Magellanic Cloud (LMC), and it was the first extragalactic pulsar
discovered \citep{mha+83}.  After its discovery, the pulsar was
studied at frequencies near 600 MHz by \citet{mmh+91} and
\citet{cmh91}, and timing results for the pulsar were subsequently
published by \citet{ckm+01}.

PSR J0529$-$6652 was redetected as a known pulsar in a more recent
1400 MHz Parkes survey of the Magellanic Clouds (MCs) for pulsars
\citep{ckm+01, mfl+06}. In these redetections, it was evident that the
pulsar was variable on a time-scale of minutes or less and showed
possible nulling behavior, unlike any of the other MC radio pulsars
detected. PSR J0529$-$6652 also had clearly detectable single radio
pulses in these observations, making this pulsar phenomenologically
unlike every other extragalactic radio pulsar yet
discovered.\footnote{The LMC pulsar PSR B0540$-$69 also has detectable
single radio pulses \citep{jr03}, but it is not detectable as a
periodic source at 1400 MHz. It is also a young, rapidly rotating
X-ray emitter \citep{shh84}, making it different from PSR J0529$-$6652
in this respect as well.}  PSR J0529$-$6652 is very luminous at 1400
MHz, ranking in the top 2\% of the known radio pulsar population in
the ATNF pulsar catalog \citep{mht+05}. However, apart from its
luminosity and apparent nulling behavior, the pulsar is unremarkable,
with a period and period derivative that are typical of the unrecycled
radio pulsar population.

Previous measurements of PSR J0529$-$6652 taken with Parkes at 600 MHz
by \citet{cmh91} showed that the pulsar has a simple, single-peaked
morphology, with little pulse broadening.  The width of the measured
profile is $\sim 3$\% of the period at 600 MHz \citep{cmh91}.  The
integrated pulse profile of PSR J0529$-$6652 at 1400 MHz is also
single-peaked and uncomplicated (see our analysis below), and its
width at both frequencies is consistent with the empirical relation
for the pulse width at 1 GHz of $W = 2.45^{\circ} P^{-1/2} / \sin
\alpha$ presented by \citet{r90} for core emitters, if a reasonable
magnetic inclination angle is assumed ($\alpha \sim
15^{\circ}$). These features are consistent with a core emission
interpretation for PSR J0529$-$6652.  A polarization profile at 600
MHz from \citet{cmh91} shows significant linear polarization ($\sim
25$\%), with weaker circular polarization that changes sign
symmetrically near the profile peak. This sign reversal also supports
the interpretation of PSR J0529$-$6652 as a single core emitter
\citep{r90}. In addition, the majority of known pulsars ($\sim 70$\%;
Rankin 1990\nocite{r90}) that have been classified have core
components.  On this basis, we conclude that PSR J0529$-$6652 is
likely to be exhibiting core emission, making it useful for a test of
the \citet{gs00} model.

\section{Observations and Analysis}

In order to investigate the pulse variability characteristics of PSR
J0529$-$6652, we observed the pulsar with the Parkes 64-m radio
telescope in two separate observations of length 5.0 and 3.8 hr in Feb
2008.  Both observations were conducted at a center frequency of 1390
MHz using the center beam of the multibeam receiver and analog
filterbank system \citep{swb+96, mlc+01}. A bandwidth of 256 MHz was
split into $512 \times 0.5$ MHz channels, and each channel was 1-bit
digitized and sampled every 500 $\mu$s (see Table \ref{tbl-1}).  The
observing setup was identical to the one used in a radio pulsar search
of the X-ray binary XTE~J0103$-$728 in the Small Magellanic Cloud
\citep{cld+09}, which was part of the same observing campaign.  Radio
frequency interference (RFI) can be a significant problem at 1400 MHz
at Parkes, and much of the data for PSR J0529$-$6652 was at least
partially corrupted by RFI. We selected a portion of the first
observation that was clean of RFI for the analysis. This subset
consisted of 4195.5 seconds of integration, corresponding to 4299
complete pulses.

\begin{figure}
\centerline{\psfig{figure=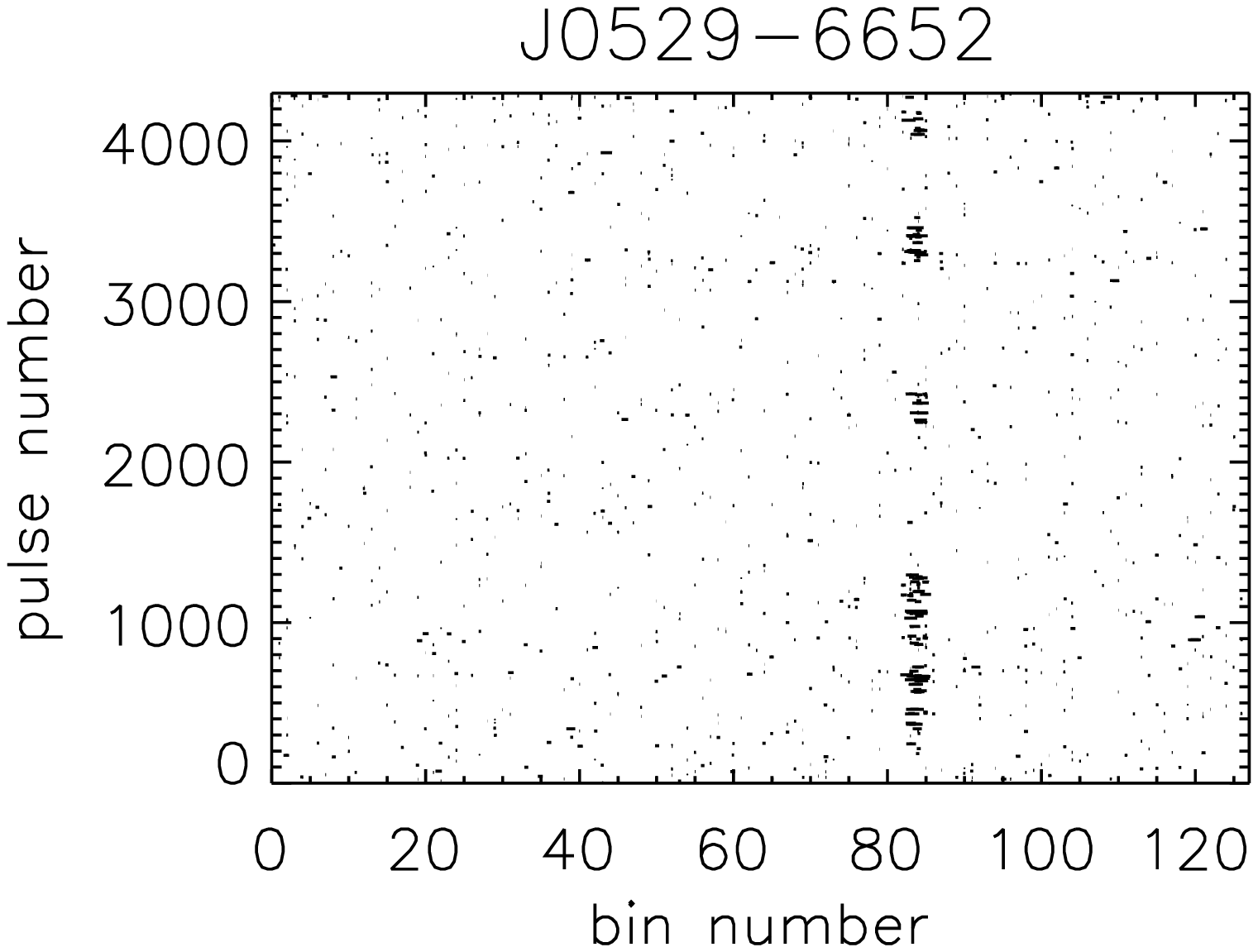,width=3.5in,angle=0}
\psfig{figure=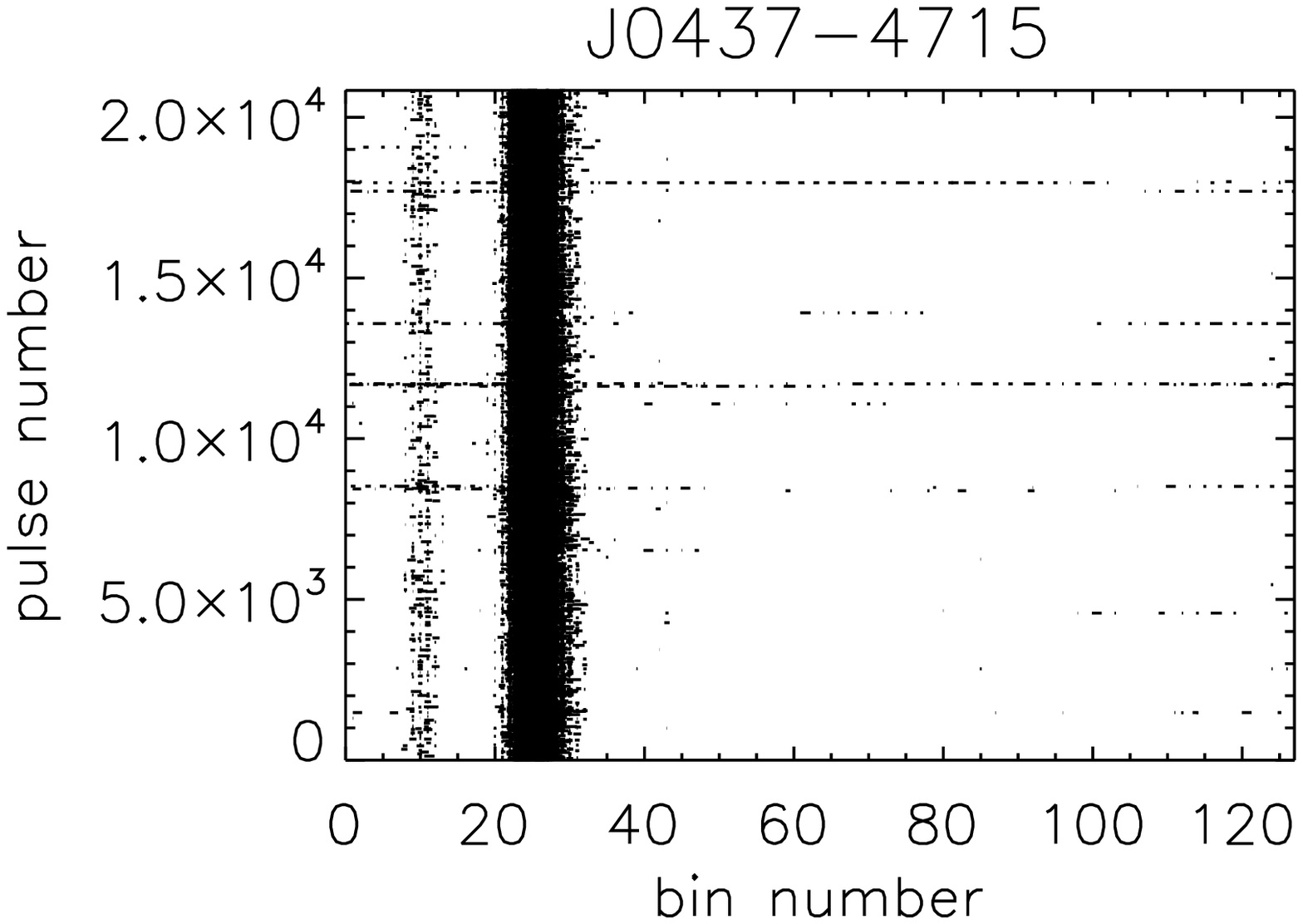,width=3.5in,angle=0}}
\centerline{\psfig{figure=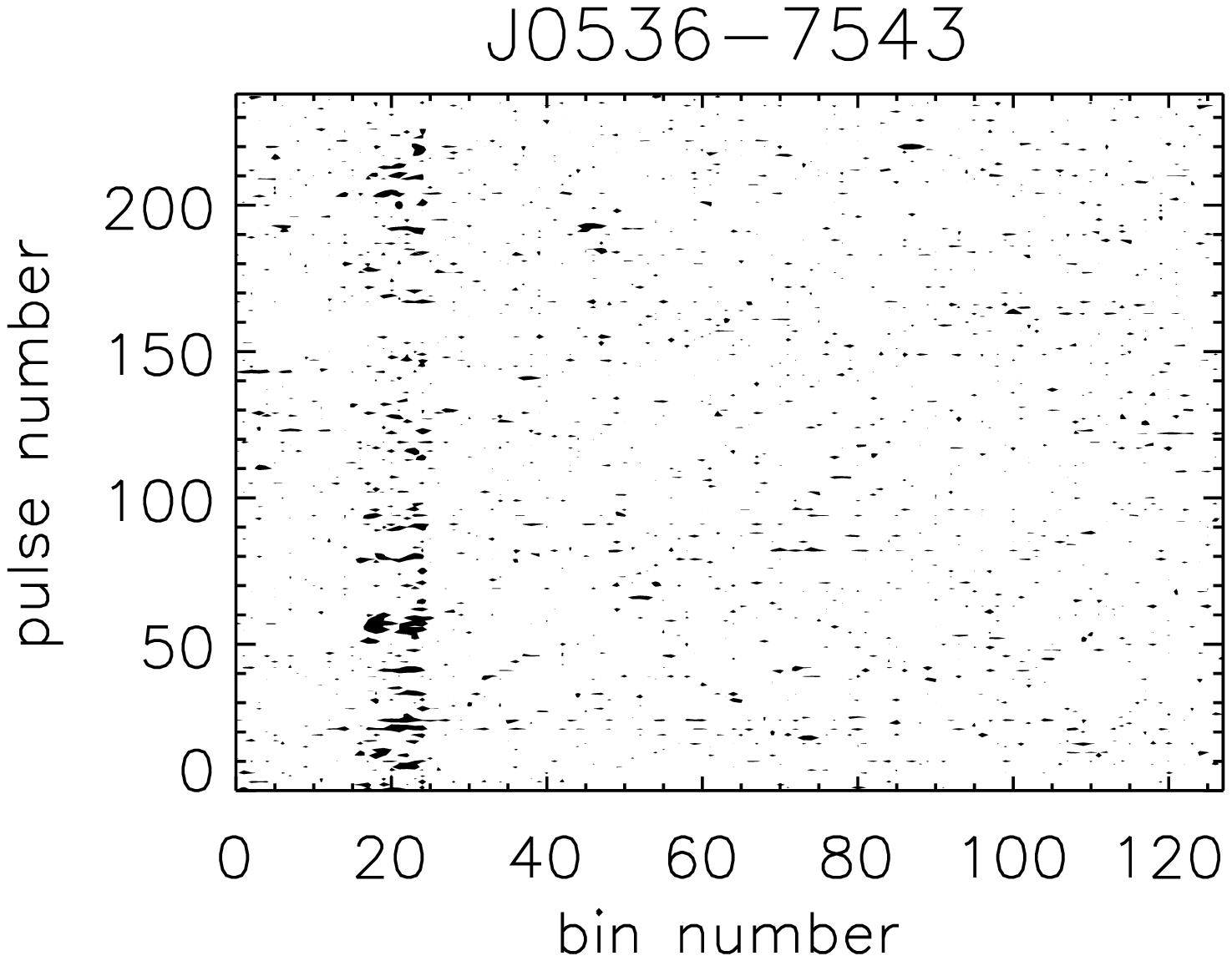,width=3.5in,angle=0}
\psfig{figure=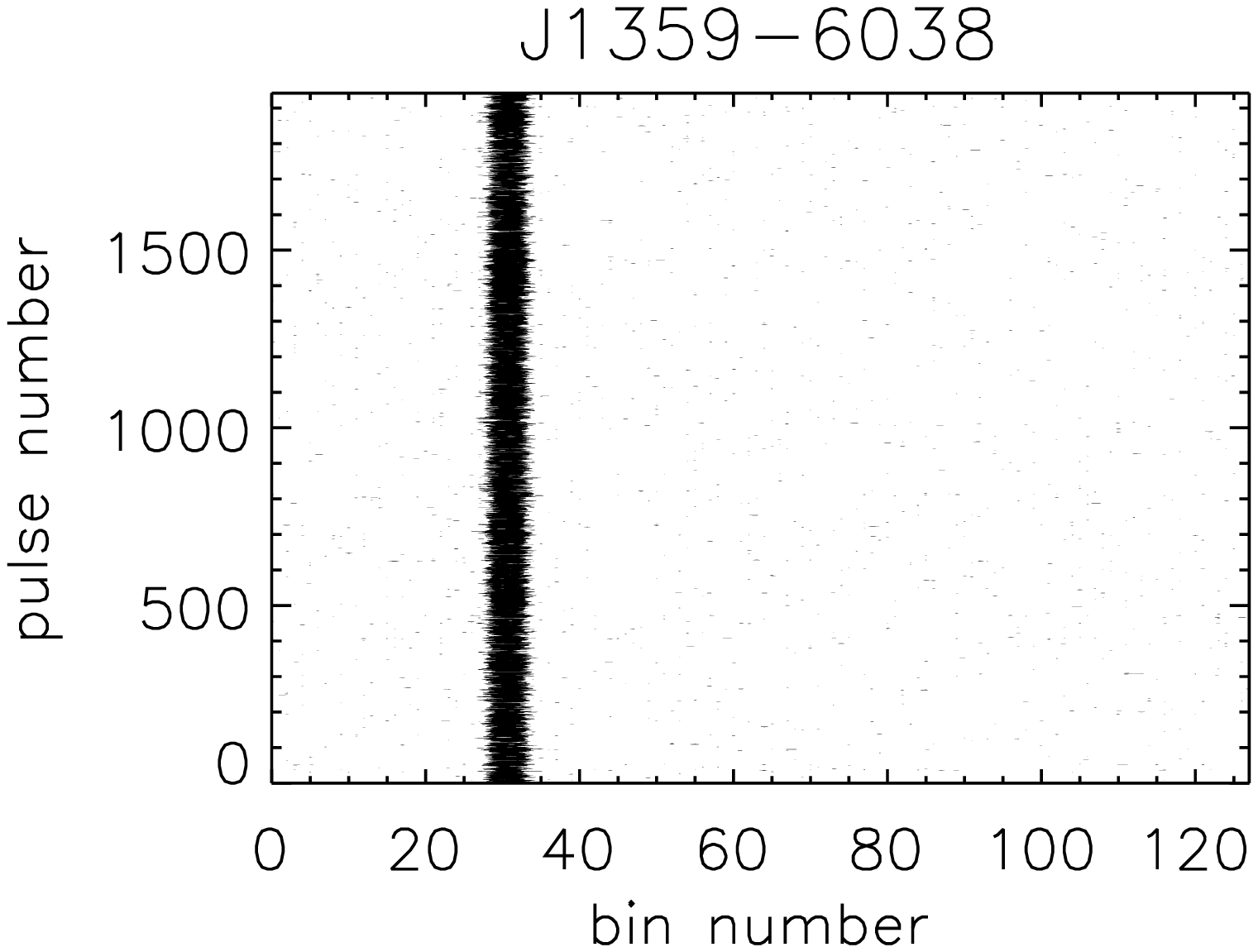,width=3.5in,angle=0}}
\caption{Pulse stacks for PSR J0529$-$6652 and three bright test
pulsars (PSRs J0437$-$4715, J0536$-$7543, and J1359$-$6038).  Each
pulse stack has 128 pulse phase bins (horizontal axis) but a different
number of consecutive pulses (vertical axis). Table \ref{tbl-1} lists
the observing parameters and ATNF catalog properties for the four
pulsars.  The contrast in each plot has been adjusted to best
illuminate any variability in the pulses.  All observations were taken
near 1400 MHz, and the data are largely free of RFI. The variability
is clearly evident in the PSR J0529$-$6652 pulse stack, consistent
with its large measured NF (Table \ref{tbl-1}).\label{fig-1}}
\end{figure}

\begin{figure}
\centerline{\psfig{figure=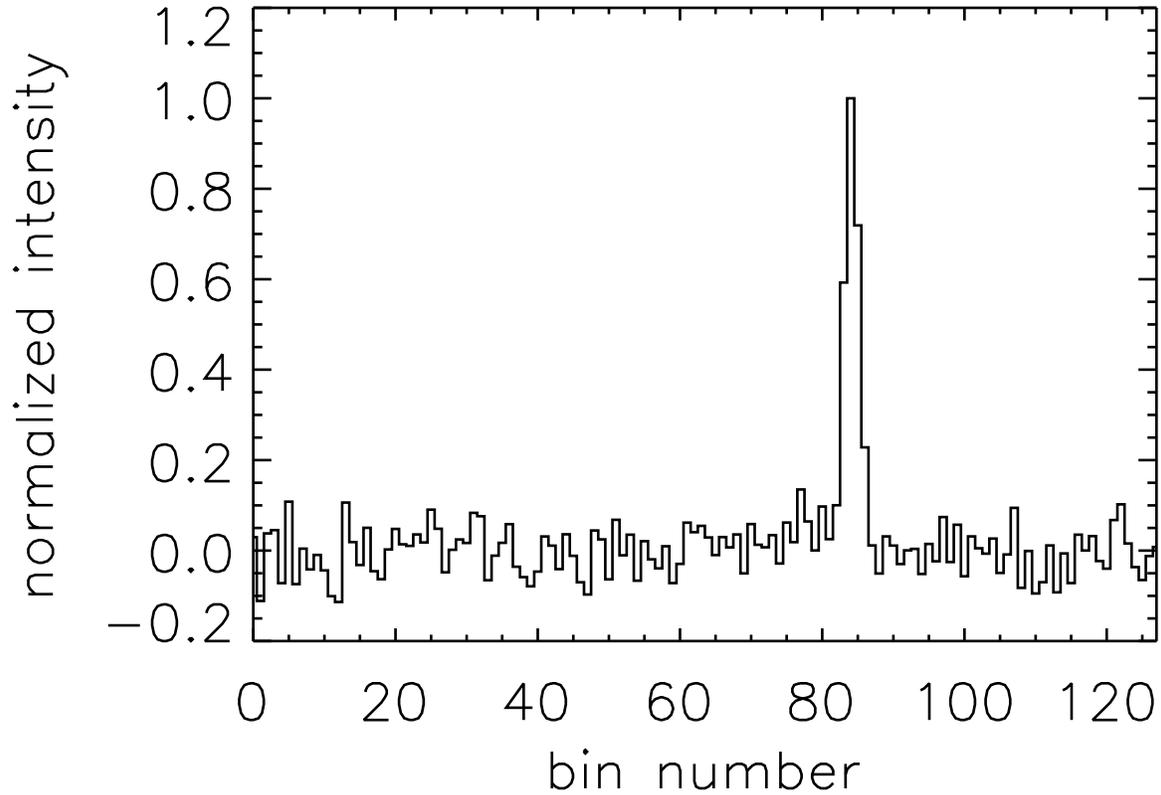,width=7.0in,angle=0}}
\caption{Normalized mean intensity profile for PSR J0529$-$6652 from
the addition of 4299 consecutive pulses from a 1390 MHz Parkes
observation. There are 128 phase bins in the profile, and the profile
has unity peak value and an off-pulse mean of zero.  The mean pulse
profile is narrow and uncomplicated, with a width of 3-4 bins ($\sim
25$ ms, or $\sim 3$\% of the pulse period), and has no obvious
additional or outlying components. These features, plus the
polarization characteristics measured at 600 MHz by \citet{cmh91},
suggest that PSR J0529$-$6652 is likely to be exhibiting core
emission.\label{fig-2}}
\end{figure}

We performed the following operations on PSR J0529$-$6652 as well as
on three bright test pulsars, PSRs J0437$-$4715, J0536$-$7543, and
J1359$-$6038, in order to test our processing algorithms.\footnote{All
three test pulsars were observed with the same Parkes observing system
as PSR J0529$-$6652 (but at different epochs). In the case of PSR
J1359$-$6038, a slightly different center frequency (1374 MHz) and
filterbank system (consisting of $96 \times 3$ MHz channels) were used
\citep{mlc+01} (see Table \ref{tbl-1}).} We dedispersed the raw
channelized data for each pulsar using the catalog dispersion measure
(DM) from the ATNF pulsar catalog
\citep{mht+05}\footnote{http://www.atnf.csiro.au/research/pulsar/psrcat/}
to produce a dedispersed time series.  We then created a pulse stack
(e.g., Weltevrede et al. 2006a\nocite{wes06}; Burke-Spolaor et
al. 2012\nocite{bjb+12}) by folding the dedispersed time series modulo
the topocentric pulse period and stacking the resulting consecutive
pulses. This produced a two-dimensional array consisting of pulse
number vs. pulse phase bin. In each case, 128 pulse phase bins were
used for the pulse stack.  We determined the mean pulse profile by
summing the pulses in the pulse stack while preserving pulse
phase. All elements in the pulse stack were then subsequently adjusted
(normalized) so that the calculated mean intensity had zero mean for
the off-pulse bins and unity peak value.  The mean intensity and mean
intensity squared were then recomputed for each bin using this
normalized pulse stack.  Fig. \ref{fig-1} shows the normalized pulse
stack for J0529$-$6652 and the three test pulsars for comparison, and
Fig. \ref{fig-2} shows the mean intensity profile for PSR
J0529$-$6652. As is the case at 600 MHz, the mean pulse profile for
PSR J0529$-$6652 at 1390 MHz has a simple morphology and narrow width,
with a duty cycle of 3-4 bins (corresponding to $\sim 25$ ms, or $\sim
3$\% of the pulse period).  This similarity to the width measured by
\citet{cmh91} at 600 MHz indicates that the pulse does not experience
significant broadening at lower frequencies and remains narrow across
a range of frequencies. The phase-resolved modulation index was
measured for each profile phase bin using the normalized pulse stack
according to Eq. \ref{eqn-2}. An uncertainty in each modulation index
value was computed using the rms values of the intensity and intensity
squared within each bin and the number of points that were summed
within each bin. The modulation index values were corrected to account
for contributions from the interstellar medium (ISM) (this is
discussed in Section 3.2), and the resulting phase-resolved modulation
index for PSR J0529$-$6652 is plotted and discussed later in the
paper.

Nulling fractions (NFs) were also computed for PSR J0529$-$6652 and
the three test pulsars using on and off-pulse intensity histograms
that were created from the pulse stacks, as outlined by
\citet{wmj07}. Prior to this, however, we tested our procedure by
recomputing the NF for three known nulling pulsars measured at Parkes
by \citet{wmj07} at a center frequency of 1518 MHz. These pulsars were
PSRs J1049$-$5833, J1502$-$5653, and J1525$-$5417 (see their Table
1). We compared our NF values as a check for consistency.  Two of the
archival Parkes observations that we used for this comparison were
also taken at 1518 MHz, while the third (PSR J1049$-$5833) had a
center frequency of 1318 MHz.  The uncertainties in our measured NFs
were estimated using $N^{-1/2}$, where $N$ is the number of
subintegrations (pulses) used. This is a slightly more conservative
estimate than that used by \citet{wmj07}, where the number of null
subintegrations $n_{p}$ and total number of subintegrations $N$ were
used in the expression $\sqrt{n_{p}} / N$.  In all three test cases we
obtained values that were consistent with the \citet{wmj07} values
considering our respective uncertainties and that different data sets
were used. This NF comparison is presented in Table \ref{tbl-2}. As
mentioned below, the NF measured for PSR J0529$-$6652 may be
influenced by the much larger distance to the pulsar than to the test
pulsars, possibly making many of the pulses from PSR J0529$-$6652 unseen.

\section{Results and Discussion}

\subsection{Giant Pulses from PSR J0529$-$6652 and the Nulling Fraction}

\begin{figure}
\centerline{\psfig{figure=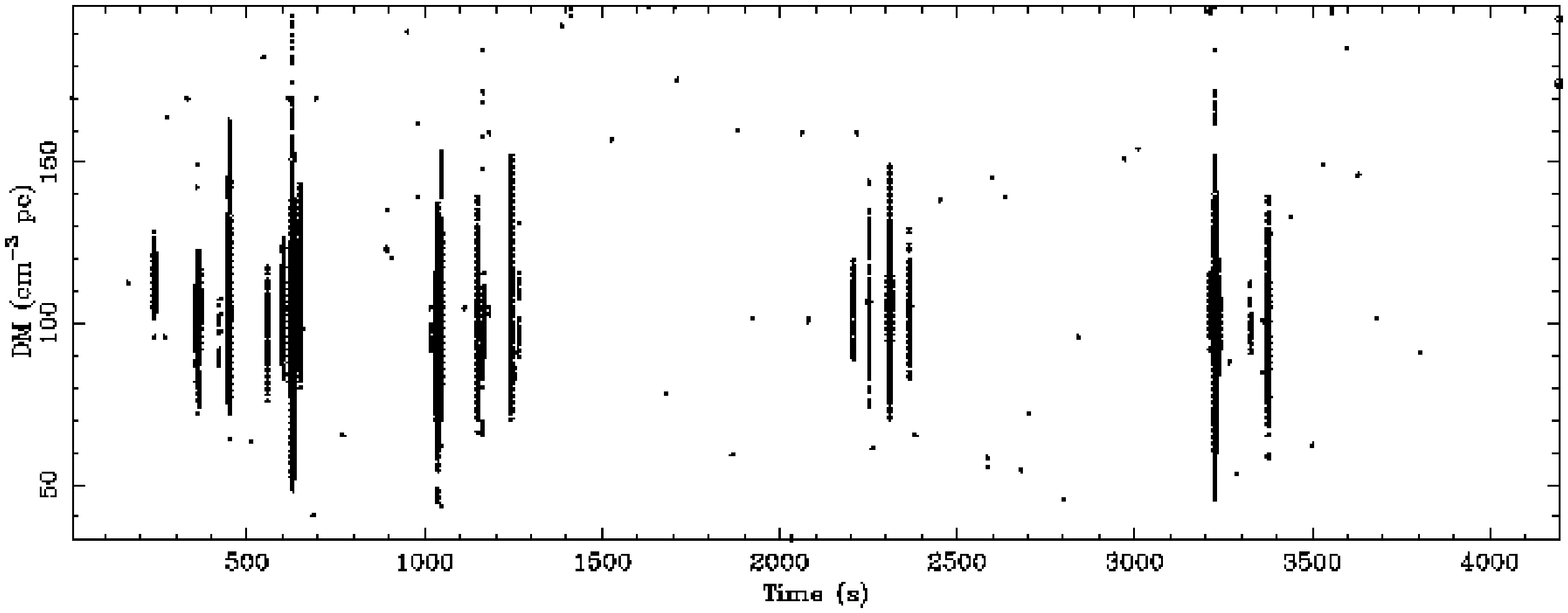,width=7.4in,angle=270}}
\centerline{\psfig{figure=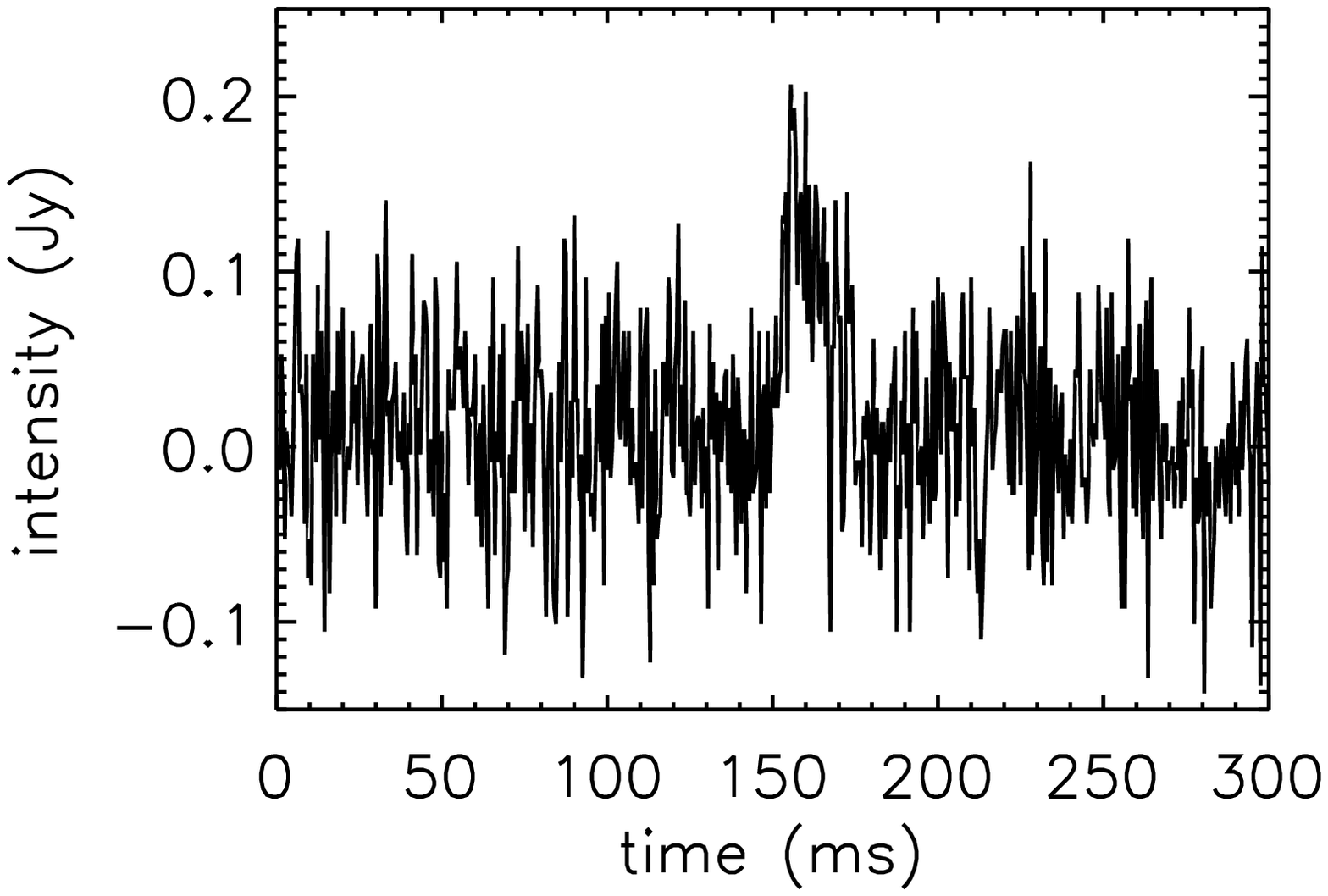,width=2.3in,angle=0}
            \psfig{figure=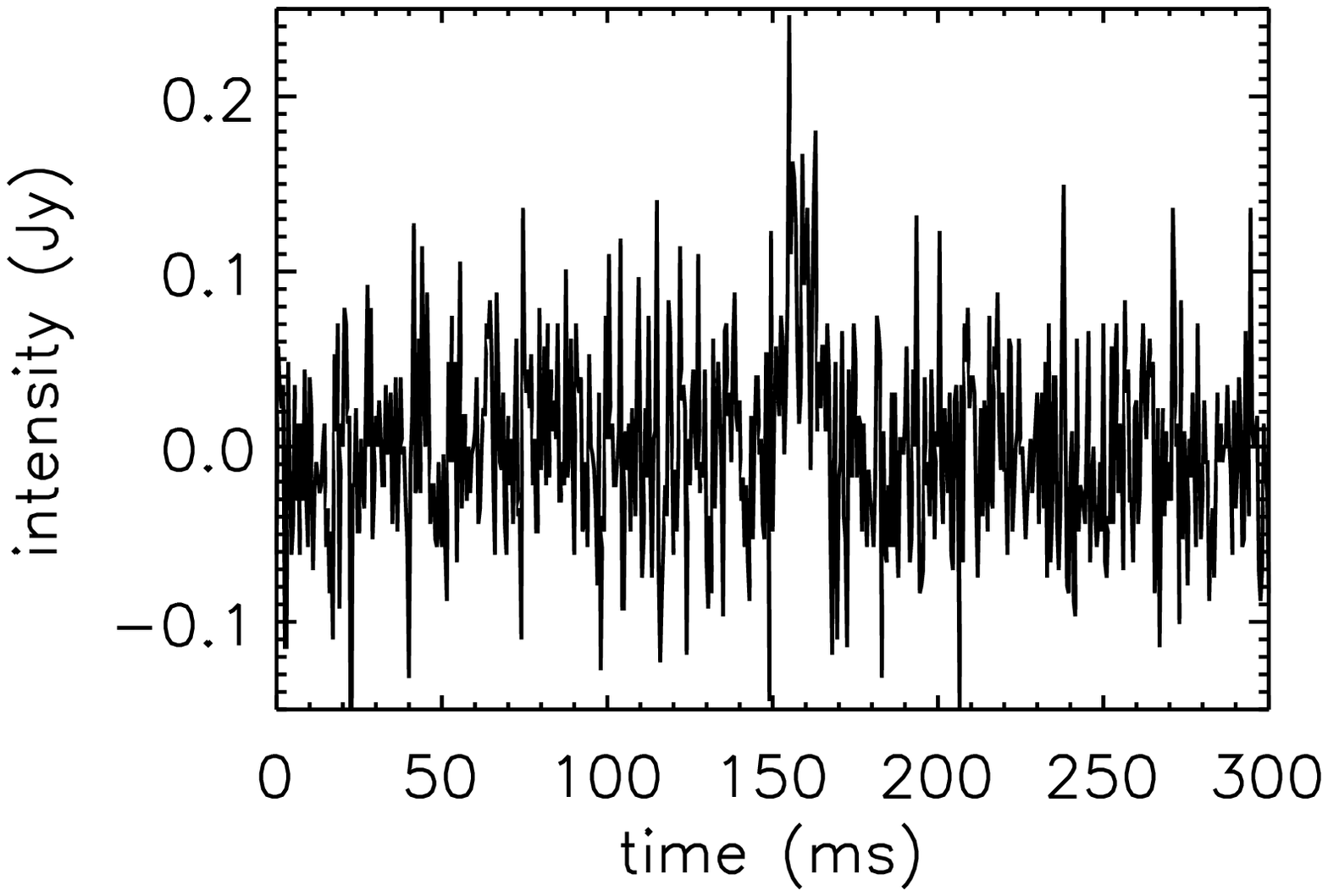,width=2.3in,angle=0}
            \psfig{figure=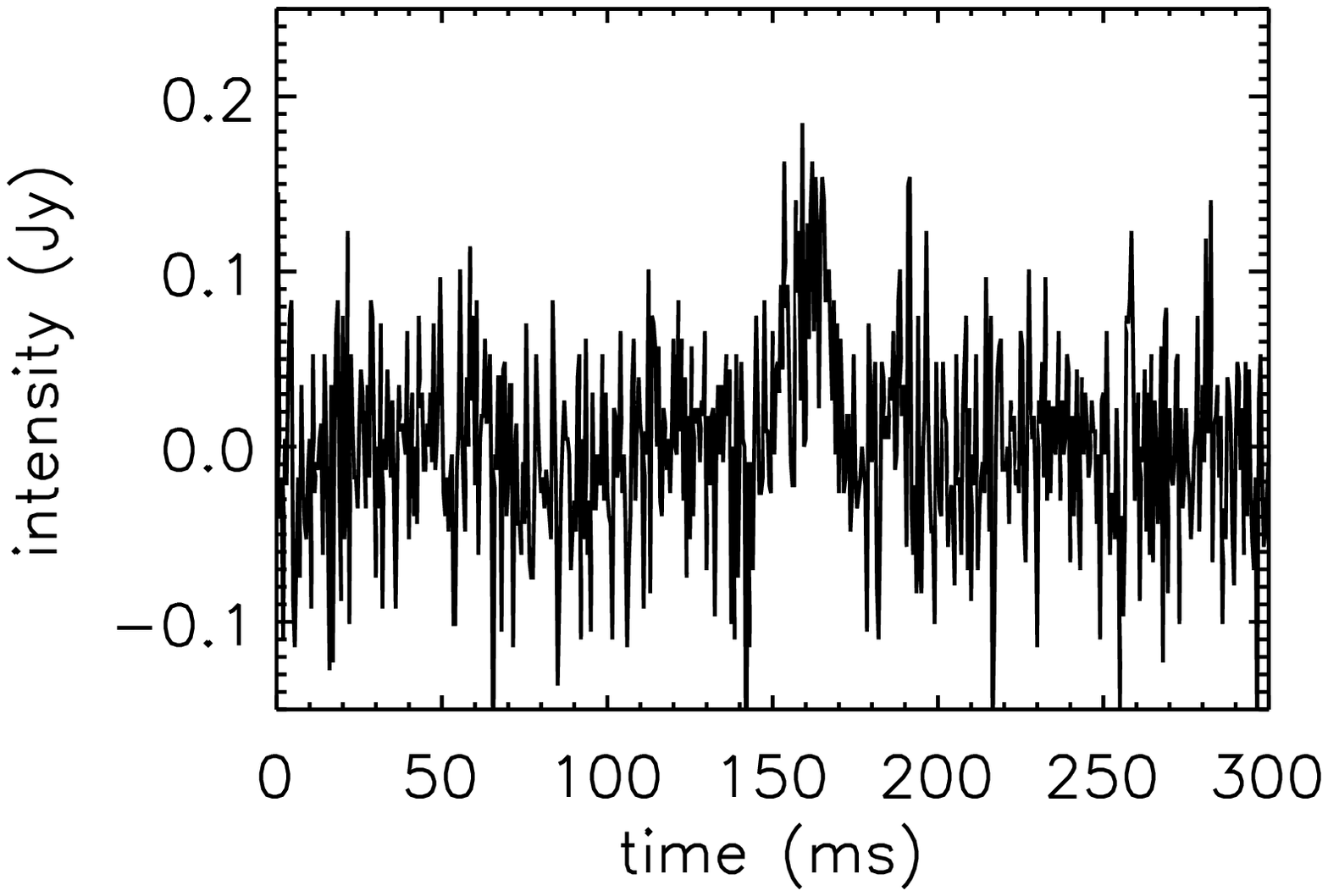,width=2.3in,angle=0}}
\caption{Single pulse detections of PSR J0529$-$6652 at 1390 MHz from
  4195.5 s of integration time, or 4299 pulses.  The data here are the
  same data that were used to make the pulse stack and to measure the
  modulation index and NF (see Table \ref{tbl-1}). The top frame shows
  pulse strength as a function of both DM and time, with pulse events
  with S/N $> 5.5$ shown.  As expected, the pulses occur most strongly
  near the pulsar's DM of $\sim 100$ pc cm$^{-3}$.  The lower frames
  show three detectable pulses from this observation plotted as
  intensity vs. time after dedispersion was applied.  In all three
  cases, the pulses have a width of $\sim 20$ ms ($\sim 2$\% of the
  pulse period), or roughly 40 samples. This is comparable to the
  width of the integrated pulse profile (see Figs. \ref{fig-2} and
  \ref{fig-3}). Since the pulses are not dispersion or scatter
  broadened, this suggests that we are seeing the intrinsic widths and
  that the pulses are not giant micropulses.  These three pulses also
  occur at the same pulse phase as the integrated profile.  PSR
  J0529$-$6652 is the second pulsar in the LMC (after PSR B0540$-$69)
  known to emit detectable single radio pulses. The pulses shown here
  illustrate the high degree of amplitude variability for the pulsar,
  which is confirmed by the large measured modulation index and
  NF.\label{fig-5}}
\end{figure}

Individual pulses from PSR J0529$-$6652 were clearly detected in the
4195.5 s subset of the observation used in the analysis
(Fig. \ref{fig-5}).  Three of these pulses, shown in the lower part of
Fig. \ref{fig-5}, were constructed using the dedispersed data. In
these three plots, the intensity (shown on the vertical axis) was
calculated by equating the rms of the dedispersed time series to the
noise level as determined by the radiometer equation (see, e.g., Table
1 of Burke-Spolaor et al. 2011\nocite{bbe+11}).  The pulse flux
density was calculated from the rms and the pulse signal-to-noise
ratio (S/N), which ranged from 8 to 11 for these three pulses. The
resulting flux densities ranged from $S \sim 420$ to 575 mJy at 1400
MHz. Using the definition of the pseudoluminosity (e.g., Lorimer \&
Kramer 2005\nocite{lk05}), $L = S d^{2}$, and assuming a LMC distance
of $d \sim 50$ kpc \citep{k09}, this corresponds to 1400 MHz peak
luminosities of between 1050 and 1440 Jy kpc$^{2}$.  All three pulses
have widths of $\sim 20$ ms, corresponding to about 40 time samples,
which is comparable to the width of the integrated pulse profile shown
in Fig. \ref{fig-2}. This is much larger than either the dispersion
smearing within the frequency channels ($\sim 0.15$ ms) or the
expected pulse scattering time ($\sim 10^{-4}$ ms) from Galactic
plasma at 1400 MHz, as estimated from the NE2001 electron model
\citep{cl02}.  This indicates that the pulses are not giant
micropulses or very narrow single pulses like those seen for the Crab
and Vela pulsars and for PSR B1937+21 \citep{cst+96,jvk+01,jr04}.  The
three pulses shown also occur at the same pulse phase as the peak of
the integrated profile.

\begin{figure}
\centerline{\psfig{figure=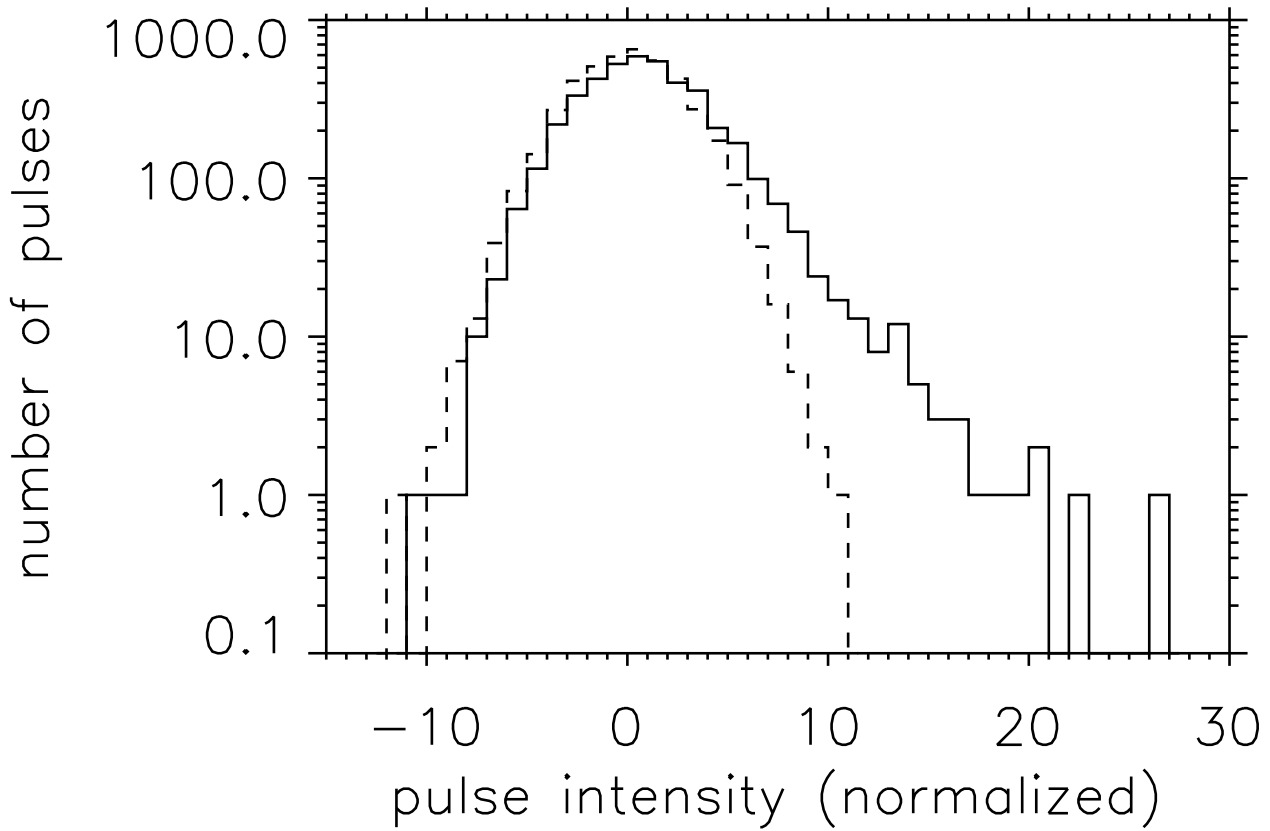,width=7.0in,angle=0}}
\caption{Histogram of on-pulse intensities for PSR J0529$-$6652 from
the subset of 4299 pulses used in the analysis (solid line). The
corresponding off-pulse intensities calculated using the same number
of off-pulse phase bins for each pulse are also shown (dashed
line). Both histograms have been normalized to the mean on-pulse
intensity value. These histograms were used in the calculation of the
NF for PSR J0529$-$6652.  There is an excess of pulses with large
amplitudes, extending well beyond the noise limit, indicating that PSR
J0529$-$6652 is a giant pulse emitter. We are unable to distinguish
between a power-law and lognormal distribution for the giant pulses
owing to the relatively small number of pulses
detected.\label{fig-histogram}}
\end{figure}

A comparison of pulse intensities in the sample to the average pulse
intensity indicates that some pulses have intensities $\ga 20$ times
the average. This can be seen in Fig. \ref{fig-histogram}, which shows
the pulse intensity histogram.  These pulses may be considered giant
pulses, which are generally defined as having amplitudes $> 10$ times
the average \citep{jvk+01, k06, kss11}.  However, it is not clear from
the small number of high-intensity pulses shown in the histogram in
Fig. \ref{fig-histogram} whether the giant pulses follow a power-law
distribution in intensity, as expected for classical giant pulses
(e.g., Argyle \& Gower 1972\nocite{ag72}), or possibly a lognormal
distribution, as in the case of PSR B0656+14 (Weltevrede et
al. 2006b\nocite{wsr+06}; see discussion below). A fit to the data
is unable to
distinguish between these two cases. Future observations should
provide better statistics and should be able to resolve this question.  There
is also no evidence of any association with a high-energy emission
mechanism for PSR J0529$-$6652, which is thought to be a feature of
classical giant pulses \citep{jr04}. We checked this with a
high-resolution {\it XMM-Newton} map of the region (Filipovi{\'c},
priv. comm.; see also Bozzetto et al. 2012\nocite{bfc+12}) and by
folding {\it Fermi} Large Area Telescope (LAT) data using the pulsar's
ephemeris to look for gamma-ray pulsations. No emission
was detected from the pulsar in either case.

Another test for the presence of giant pulses is an R-parameter,
defined by \citet{jvk+01} and described by \citet{bjb+12}.  A value
$R$ is determined for each profile bin by taking the difference
between the maximum value and the mean value of the samples for that
bin, which is then divided by the rms of the values for that bin. A
subsequent comparison of the R-value for each bin with the rms of the
R-values for the off-pulse bins can be used as a threshold test for
the presence of sparse modulated emission, such as giant
pulses. Following \citet{bjb+12}, we use a significance threshold
defined as $R$ minus the off-pulse mean of the $R$ values, divided by
the standard deviation of $R$ in the off-pulse window. If this
quantity is greater than 4 for any phase bin, then giant pulses are
considered to be present. For three of the four on-pulse bins for PSR
J0529$-$6652, $R$ was larger than 6 and nowhere else in the profile
was it above 4. This supports our conclusion that giant pulses from
PSR J0529$-$6652 are present and that they do not significantly lead
or trail the main pulse.

As seen in Fig. \ref{fig-5}, there are at least 15 distinct pulses
that are visible by eye that were also detected in the single pulse
search detection algorithm \citep{cm03}.  This corresponds to one
detectable pulse every $\sim 4.5$ min on average, though it is evident
that the pulses are not evenly spaced. For comparison, we consider two
young, canonical giant radio pulse emitters: the Crab pulsar and PSR
B0540$-$69.  \citet{cmj+05} used the same Parkes observing system as
used here (but with a different sampling rate) to observe the X-ray
pulsar PSR J0537$-$6910 in the LMC at 1400 MHz \citep{mgz+98}.  They
calculated that if the Crab pulsar were located in the LMC, one pulse
would be expected to be detectable with this system every $\sim 20$
min. In the case of PSR B0540$-$69, which is in the LMC, one giant
pulse should be detectable every $\sim 30$ min.\footnote{This expected
detection rate for PSR B0540$-$69 is somewhat larger than the actual
detection of only one pulse in a recent 1.4 hr test observation that
had comparable sensitivity. This observation was taken as part of new
multibeam pulsar survey of the LMC with Parkes.}  Thus, the rate of
detectable pulses per unit time from PSR J0529$-$6652 with this system
is almost an order of magnitude greater than the expected rates for
either the Crab pulsar (if it were in the LMC) or PSR
B0540$-$69. Moreover, the spin period of PSR J0529$-$6652 (976 ms) is
more than an order of magnitude larger than either the Crab pulsar (33
ms) or PSR B0540$-$69 (50 ms). Therefore, the likelihood that any one
pulse from PSR J0529$-$6652 would be detectable as a giant pulse with
this system (or, alternatively, the fraction of emitted pulses
detectable as single pulses with this system) is about two orders of
magnitude greater than for either the Crab (if it were in the LMC) or
PSR B0540$-$69.  Unlike these two pulsars, however, PSR J0529$-$6652
is physically unremarkable, making this feature of its emission quite
unusual and unexpected.

The pulse characteristics from PSR J0529$-$6652 appear to be more
similar to those from PSR B0656+14 \citep{wsr+06,teh+12}. The pulses
from PSR B0656+14 follow a lognormal distribution rather than a 
power-law (e.g., Fig. 1 of \citet{wsr+06}, which might also be the case for
PSR J0529$-$6652 (see Fig. \ref{fig-histogram}), they are not
associated with any high-energy emission, and they are not narrow
micropulses.  PSR B0656+14, like PSR J0529$-$6652, also does not have
a large magnetic field at the light-cylinder radius, which is a
feature of known classical giant pulse emitters \citep{cst+96}.  These
features make both PSR J0529$-$6652 and PSR B0656+14 different from
classical giant pulse emitters.  As discussed below, the large
modulation indices measured for PSRs B0656+14 and J0529$-$6652 and the
proximity of both pulsars to the bulk of the unrecycled pulsar
population on the period/period derivative diagram also suggest a
similarity between these two sources.

We measured NFs for PSR J0529$-$6652 and our three test pulsars. These
values are presented in Table \ref{tbl-1}. Both PSRs J0437$-$4715 and
J1359$-$6038 have essentially no nulling near 1400 MHz (NF $\sim
0$). The NF for PSR J0529$-$6652 at 1390 MHz is $83.3 \pm 1.5$\%,
indicating that either radio emission is not present for the majority
of the pulse periods or that most pulses are simply too weak to be
distinguishable from noise. This latter point is an important
possibility given the very large distance to PSR J0529$-$6652. In all
four cases the measured NFs appear qualitatively consistent with the
pulse stacks shown in Fig. \ref{fig-1}.

\subsection{The Modulation Index of PSR J0529$-$6652}

For the modulation index calculation, we first estimated the
contribution to the modulation from the Galactic ISM for each of the
pulsars, and we corrected the measured modulation indices for this to
obtain the value intrinsic to each pulsar.  As outlined in
\citet{jg03}, the measured ($m$), intrinsic ($m_{i}$), and
ISM-produced ($m_{ISM}$) modulation indices are related according to:

\begin{equation}
(m^{2} + 1) = (m_{i}^{2} + 1)(m_{ISM}^{2} + 1).
\end{equation}

$m_{ISM}$ can be estimated using the relationship \citep{jap01, jg03}

\begin{equation}
m_{ISM} = \frac{1}{\left( 1 + \eta B / \delta \nu \right)^{1/2}},  
\end{equation}

where $B$ is the observing bandwidth, $\delta \nu$ is the
characteristic ISM diffractive scintillation bandwidth, and $\eta$ is
a coefficient ranging from 0.1 to 0.2 \citep{cwd+90}. $\delta \nu$ was
determined at 1400 MHz for each pulsar using the NE2001 model of
\citet{cl02} and the pulsar's catalog DM. For PSR J0529$-$6652, we
used the maximum Galactic DM contribution along its line of sight
according to the NE2001 model (52 pc cm$^{-3}$).  The uncertainty in
$m _{ISM}$ was calculated using a range of $\eta$ of 0.1 to 0.2.
Following the method of \citet{jg03}, the uncertainty in the intrinsic
modulation index $m_{i}$ was determined by using the larger of the
measurement uncertainties in $m$ and $m_{ISM}$. The results are
presented in Table \ref{tbl-1}.

It should be noted that in the case of PSR J0529$-$6652, $m_{ISM}$
does not include any contribution from propagation through the LMC,
where the pulsar resides. There is no model that we can use to
estimate the LMC contribution to the modulation, but observations of
other known pulsars in the MCs do not show any of the extreme
variability and null-like behavior that is seen for PSR J0529$-$6652
\citep{ckm+01,mfl+06}. We conclude from this that the contribution to
the modulation from the LMC itself is likely to be minimal and does
not significantly affect our results.

\begin{figure}
\centerline{\psfig{figure=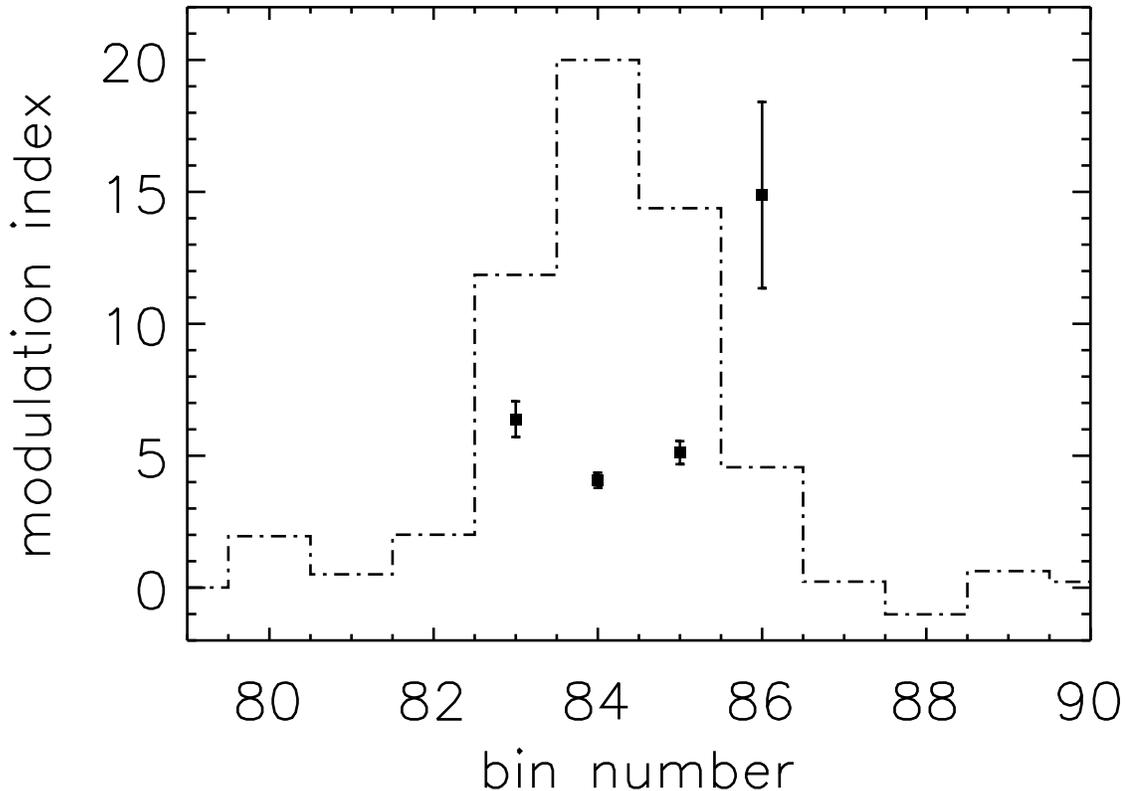,width=7.0in,angle=0}}
\caption{Phase-resolved intrinsic modulation index values measured for
PSR J0529$-$6652 (squares) overlaid with corresponding mean intensity
values (dash-dotted line). The intensity values have been arbitrarily
scaled for display purposes. Only the pulse phase bins near the pulse
peak are shown, and only the on-pulse bins have enough signal for
reliable modulation index measurements. These modulation index values
have already been corrected for the estimated contribution from
fluctuations from the Galactic ISM.  The minimum and most precise
value of $m_{i} = 4.07 \pm 0.29$ is seen at the profile peak, and this
is the value used in our analysis (see, e.g., Jenet \& Gil
2003).\label{fig-3}}
\end{figure}

\begin{figure}
\centerline{\psfig{figure=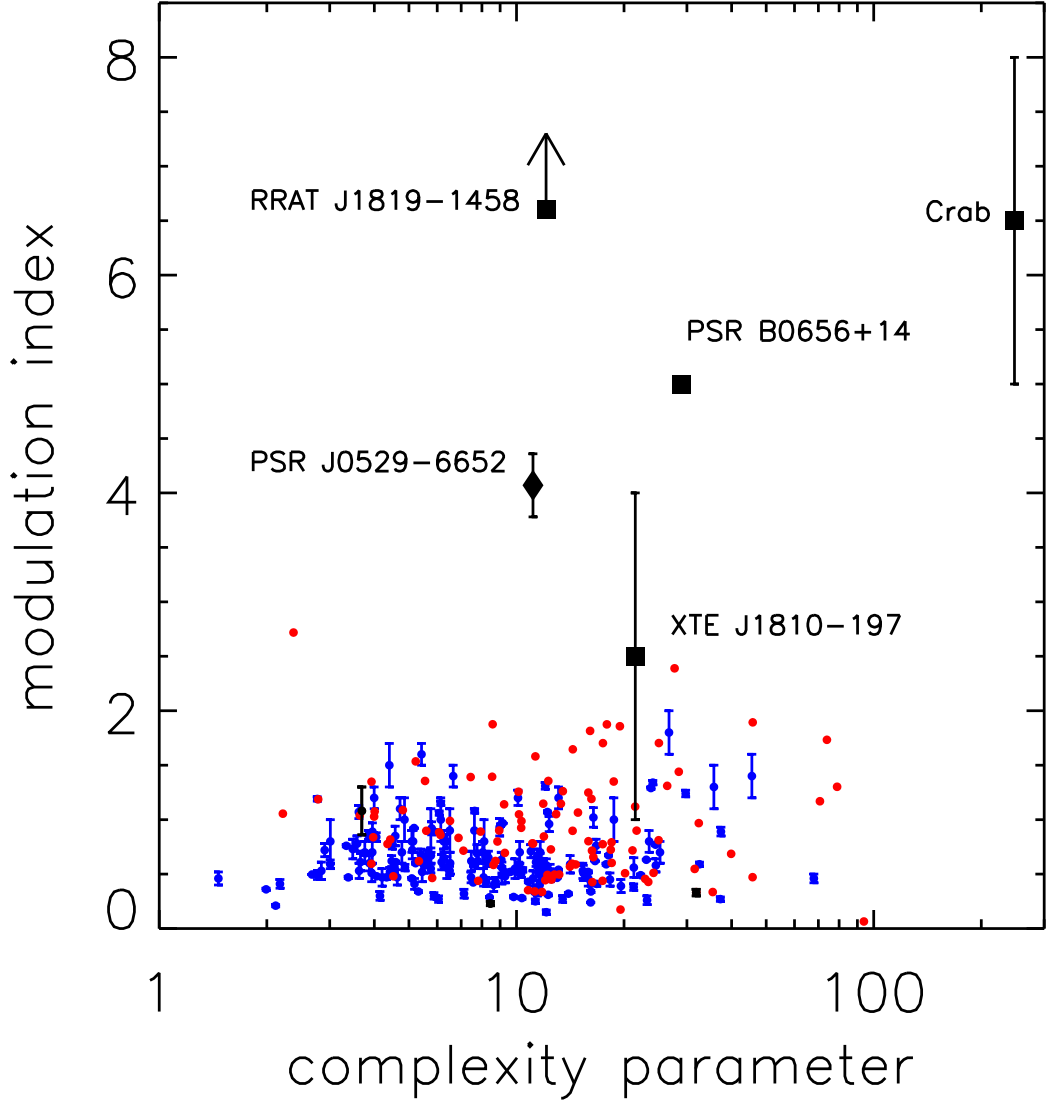,width=7.0in,angle=0}}
\caption{Intrinsic modulation index vs. complexity parameter
determined from the \citet{gs00} model is shown for several samples of
pulsars.  All of the 174 pulsars measured by \citet{wes06} (and
presented in their Table 2 and Fig. 10) are presented as small blue
circles with error bars. Also plotted are the three measurements of
our test pulsars (small black circles with errors). The 102 (out of
103) pulsars from Table 1 of \citet{bjb+12} with cataloged $\dot{P}$
values are plotted with red circles (no errors were reported).  Our
measurement for PSR J0529$-$6652 (large diamond) and the measurements
presented by \citet{wje11} in their Table 4 for four members of
different, more variable neutron star classes are also shown (large
squares).  All data shown here were taken near 1400 MHz, except for
PSR B0656+14, which was at 327 MHz.  PSR J0529$-$6652 has physical
parameters that are typical of the unrecycled radio pulsar population
(see Fig. \ref{fig-6}), unlike the other labeled neutron stars (except
possibly PSR B0656+14).\label{fig-4}}
\end{figure}

After correcting for the estimated Galactic ISM contribution, we
derived an intrinsic modulation index value for PSR J0529$-$6652 for
each profile phase bin.  Fig. \ref{fig-3} shows the modulation indices
for the on-pulse profile bins overlaid with the profile intensity.  We
selected the bin with the highest S/N (the profile peak) for the
analysis, which also had the most precise modulation index
measurement.  As stated in \citet{jg03}, this is the bin that is least
likely to be affected by outlying conal emission components, which can
introduce additional variability and can complicate the interpretation
of the modulation index as a feature of the polar cap physics. This
bin also had the lowest value of $m_{i}(\phi)$, as expected (see Jenet
\& Gil 2003\nocite{jg03}), and from this we derived an intrinsic
modulation index of $4.07 \pm 0.29$ for PSR J0529$-$6652 (Table
\ref{tbl-1}).  The modulation indices measured for our three test
pulsars using the same method are also presented in Table
\ref{tbl-1}. As expected, these values are all small ($m_{i} \la 1$)
and are consistent with the range seen for most radio pulsars (see,
e.g., Fig. \ref{fig-4}).

Fig. \ref{fig-1} shows the pulse stack for PSR J0529$-$6652, which
clearly shows its variability. As stated above, the LMC itself is
likely not a major contributor to this variability. The diffraction
scintillation time scale at 1400 MHz for PSR J0529$-$6652 from
Galactic plasma is $\sim 1500$ s \citep{cl02}, significantly longer
than the minute time scales (or less) that seem to be present
qualitatively in the pulse stack. The scintillation bandwidth is also
much smaller than the observing bandwidth (see Table \ref{tbl-1}).
This suggests that the variability of the emission is intrinsic to the
pulsar and is consistent with the large intrinsic modulation index
measured.

\begin{figure}
\centerline{\psfig{figure=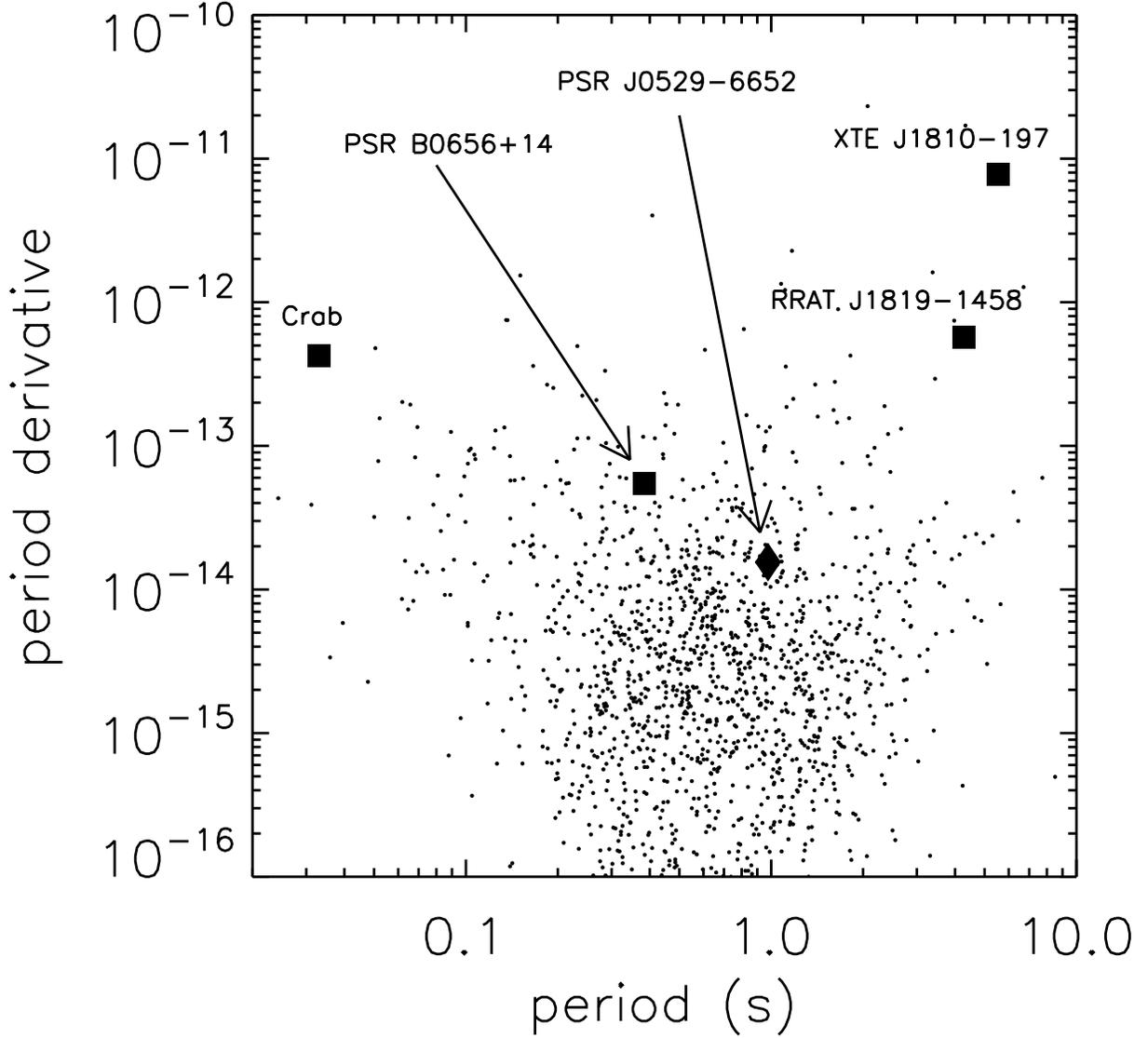,width=7.0in,angle=0}}
\caption{Period derivative vs. period for radio pulsars from the ATNF
catalog \citep{mht+05} (small dots).  Not all pulsars are shown (e.g.,
recycled millisecond pulsars are beyond the plot limits).  Also shown
are the four neutron stars with modulation index measurements
presented in Table 4 of \citet{wje11} (squares).  PSR J0529$-$6652 is
shown as the large diamond.  Compared to the other four labeled
neutron stars, PSR J0529$-$6652 lies closer to the center of the
unrecycled radio pulsar population and has spin characteristics that
are more typical of this population.\label{fig-6}}
\end{figure}

Fig. \ref{fig-4} shows the intrinsic modulation index vs. the
complexity parameter derived using the \citet{gs00} model for several
samples of pulsars.  174 pulsars measured by \citet{wes06} are plotted
as small blue dots (see their Table 2 and Fig. 10; note that their
Crab measurement was excluded in their figure and is also not included
here). In addition, 102 of the 103 pulsars measured by \citet{bjb+12}
for which there was also a cataloged $\dot{P}$ are plotted as small
red dots, but no error bars are given for these values.  Our three
test pulsars are plotted as small black dots with error bars,  
and our measurement for PSR J0529$-$6652 is indicated
by the large diamond (see also
Table \ref{tbl-1}).  All of these measurements were taken near 1400
MHz.  Although PSR J0529$-$6652 has a complexity parameter and
physical characteristics that are typical of the rest of the sample
(and the overall radio pulsar population in general; see
Fig. \ref{fig-6}), it nevertheless has a modulation index that lies
well above this sample and the sample of 12 pulsars shown in Fig. 1 of
\citet{jg03}. Only two of the pulsars measured in any of these samples
(apart from PSR J0529$-$6652) have $m > 2$.

The large modulation index for PSR J0529$-$6652 is closer to the
ranges and limits presented in Table 4 of \citet{wje11} for four
members of different neutron star classes (these are plotted in
Fig. \ref{fig-4} as large squares). These four sources are the Crab
pulsar (see above), the radio magnetar XTE J1810$-$197, the RRAT
J1819$-$1458, and the RRAT-like pulsar PSR B0656+14.  All of the
measurements of these sources were taken near 1400 MHz, except for PSR
B0656+14, which was taken at 327 MHz.  Like PSR J0529$-$6652, they all
have modulation indices that are significantly larger than the values
that have been measured for typical radio pulsars.  However, with the
possible exception of PSR B0656+14, none lies near the center of the
unrecycled radio pulsar population on the period/period derivative
diagram (Fig. \ref{fig-6}).  Thus, only PSR J0529$-$6652 (and possibly
PSR B0656+14) are typical pulsars in this sense.  In any case, the
core-like characteristics of the emission from PSR J0529$-$6652
(suggested by its simple and narrow profile morphology and its
symmetric sign change in circular polarization), plus its unremarkable
spin and physical parameters (Crawford et al. 2001\nocite{ckm+01}; see
also Fig. 6), suggest that this pulsar is more useful for a test of
the correlation predicted by the \citet{gs00} emission model than
these other neutron stars, all of which are physically unusual in
different ways.  We conclude that the large modulation index measured
for PSR J0529$-$6652 is not consistent with the \citet{gs00} emission
model, which predicts a small modulation index for a relatively large
complexity parameter.  We note also that the radio magnetar PSR
J1622$-$4950 \citep{lbb+10} has a relatively small measured modulation
index ($m \sim 0.5$ measured by \citet{bjb+12} at 1.4 GHz, and $m \sim
1.7$ measured by \citet{lbb+12} at 3.1 GHz). This indicates that not
all magnetars exhibit extreme radio modulation like XTE J1810$-$197,
just as PSR J0529$-$6652 illustrates that not all typical radio
pulsars have low modulation.

\section{Conclusions}

We have studied the variability of PSR J0529$-$6652, a luminous radio
pulsar in the LMC, using observations taken at 1390 MHz with the
Parkes 64-m telescope. The pulsar emits detectable single radio pulses
that can be classified as giant pulses, making this the second known
pulsar in the LMC (after PSR B0540$-$69) to do so.  These pulses are
comparable in width to and occur at the same phase as the integrated
profile peak, suggesting that they are not giant micropulses.  The
characteristics of the giant pulses appear to be closer to those seen
from PSR B0656+14 than from classical giant pulse emitters, such as
the Crab pulsar.  The fraction of the pulses emitted by PSR
J0529$-$6652 that are individually detectable as single pulses at this
frequency is two orders of magnitude greater than either PSR
B0540$-$69 or the Crab pulsar (if the latter were located in the LMC).
PSR J0529$-$6652 also appears to exhibit nulling behavior, and we have
measured a NF of $83.3 \pm 1.5$\% for the pulsar. Pulsed radio
emission is either not present for the majority of pulse periods, or
it is too weak in most cases to be individually detectable above the
noise.  Given the large distance to the pulsar, this is a possibility
that must be considered. Our measured intrinsic modulation index for
PSR J0529$-$6652 is $4.07 \pm 0.29$, which is significantly larger
than the values previously measured for typical radio pulsars with
similar spin and physical characteristics.  It is comparable to the
larger values presented by \citet{wje11} for several members of other
neutron star source classes that are known to be radio
variable. Unlike these other sources, however, PSR J0529$-$6652 has
spin and physical characteristics that are typical of the unrecycled
radio pulsar population. These features make this pulsar useful as a
test of the \citet{gs00} emission model, and the large modulation
index measured for PSR J0529$-$6652 does not appear to be consistent
with the model prediction.  This conclusion depends to some degree on
the assumption that PSR J0529$-$6652 is exhibiting core emission, as
seems to be the case given its simple profile morphology, narrow
profile width, and previously measured polarization characteristics at
600 MHz.

\acknowledgements 

We thank M. Filipovi{\'c} for providing us with an X-ray image of the
region containing PSR J0529$-$6652.  The Parkes radio telescope is
part of the Australia Telescope, which is funded by the Commonwealth
of Australia for operation as a National Facility managed by
CSIRO. This work was supported in part by grants from Research
Corporation, the Mount Cuba Astronomical Foundation, the National
Radio Astronomy Observatory, and the Hackman Scholarship program at
Franklin \& Marshall College.

\clearpage

\begin{deluxetable}{lcccc}
\tabletypesize{\footnotesize}
\tablecaption{Properties, Observing Parameters, and Measured Quantities for Four Pulsars\label{tbl-1}}
\tablewidth{0pt}
\tablehead{
\colhead{Pulsar} &
\colhead{J0529$-$6652} &
\colhead{J0437$-$4715} &
\colhead{J0536$-$7543} &
\colhead{J1359$-$6038} 
}
\startdata
Topocentric period, $P$ (s)\tablenotemark{a}                & 0.97573571876   & 0.00575786      & 1.24585559629   & 0.12751295       \\
Dispersion measure, DM (pc cm$^{-3}$)\tablenotemark{a}      & 103.2           & 2.6             & 17.5            & 293.7            \\
Number of consecutive pulses used, $N$                      & 4299            & 20839           & 240             & 1943             \\
Number of profile bins                                      & 128             & 128             & 128             & 128              \\
                                                            &                 &                 &                 &                  \\      
Center observing frequency (MHz)                            & 1390            & 1390            & 1390            & 1374             \\ 
Sampling time ($\mu$s)                                      & 500             & 80              & 250             & 500              \\
Observing bandwidth, $B$ (MHz)                              & 256             & 256             & 256             & 288              \\  
Scintillation bandwidth, $\delta \nu$ (MHz)\tablenotemark{b}& 1.51            & 318             & 22              & $\sim 0$         \\
                                                            &                 &                 &                 &                  \\      
Modulation index for the ISM, $m_{ISM}$\tablenotemark{c}    & $0.20 \pm 0.03$ & $0.95 \pm 0.02$ & $0.61 \pm 0.07$ & $\sim 0$         \\
Measured modulation index, $m$\tablenotemark{d}             & $4.15 \pm 0.29$ & $1.00 \pm 0.02$ & $1.40 \pm 0.22$ & $0.33 \pm 0.03$  \\
Intrinsic modulation index, $m_{i}$\tablenotemark{e}        & $4.07 \pm 0.29$ & $0.23 \pm 0.02$ & $1.08 \pm 0.22$ & $0.33 \pm 0.03$  \\
                                                            &                 &                 &                 &                  \\      
Measured nulling fraction (NF)\tablenotemark{f}             & $83.3 \pm 1.5$\%& $0.0 \pm 0.7$\% & $32.5 \pm 6.5$\%& $0.1 \pm 2.3$\%  \\ 
\enddata

\tablecomments{All observations were taken with the center beam of the
multibeam receiver and $512 \times 0.5$ MHz channel analog filterbank
system at the Parkes telescope, except for PSR J1359$-$6038, where
$96 \times 3$ MHz channels were used \citep{mlc+01}.}

\tablenotetext{a}{Obtained or derived from the ATNF pulsar catalog
\citep{mht+05}.} 

\tablenotetext{b}{Estimated from the NE2001 Galactic electron model of
\citet{cl02}.}

\tablenotetext{c}{Estimated contribution to the modulation from
propagation through the Galactic ISM.}

\tablenotetext{d}{Measured from the normalized pulse
stack. The lowest value of $m(\phi)$ among the on-pulse bins was
chosen (see \citet{jg03} and the text for justification).}

\tablenotetext{e}{Obtained after correction for the estimated Galactic ISM
contribution.}

\tablenotetext{f}{The uncertainty in NF was determined by $N^{-1/2}$,
where $N$ is the number of pulses used.}

\end{deluxetable}

\begin{deluxetable}{lccc}
\tabletypesize{\footnotesize}
\tablecaption{Nulling Fractions for Three Known Nulling Pulsars\label{tbl-2}}
\tablewidth{0pt}
\tablehead{
\colhead{Pulsar} &
\colhead{\citet{wmj07}\tablenotemark{a}} &
\colhead{This Work} &
\colhead{$N$\tablenotemark{b}} 
}
\startdata
J1049$-$5833 & $47 \pm 3$\%  & $33 \pm 35$\%     & 8    \\
J1502$-$5653 & $93 \pm 4$\%  & $70 \pm 9$\%      & 120  \\
J1525$-$5417 & $16 \pm 5$\%  & $26 \pm 5$\%      & 361  \\ 
\enddata

\tablecomments{For all three pulsars, different sets of Parkes data 
were used for our measurements than were used by \citet{wmj07}. 
All of the data used in both sets of measurements
had a center frequency of 1518 MHz, except for 
our measurement of PSR J1049$-$5833, which had a center frequency of 1318 MHz.}

\tablenotetext{a}{Values from Table 1 of \citet{wmj07}.}

\tablenotetext{b}{Number of subintegrations used in our NF
measurements. The uncertainty in the NF was determined by $N^{-1/2}$.}

\end{deluxetable}

\end{document}